# The effects of atmospheric entry heating on organic matter in interplanetary dust particles

M.E.I. Riebe[1*], D.I. Foustoukos[2], C.M.O'D Alexander[1], A. Steele[2], G.D. Cody[2], B. Mysen[2], L.R. Nittler[1]

[1]Department of Terrestrial Magnetism, Carnegie Institution of Washington, 5241 Broad Branch Road, Washington DC 20015, USA

[2]Geophysical Laboratory, Carnegie Institution of Washington, 5251 Broad Branch Road, Washington, DC 20015, USA.

*corresponding author, current address: Institute of Geochemistry and Petrology, ETH Zürich, CH-8092 Zürich, Switzerland; my.riebe@erdw.ethz.ch

Abstract

Interplanetary dust particles (IDPs) were likely major sources of extraterrestrial organics at the surface of the early Earth. However, IDPs experience heating to >500 °C for up to several seconds during atmospheric entry. In this study, we aim to understand the effects of atmospheric entry heating on the dominant organic component in IDPs by conducting flash heating experiments (4 s to 400 °C, 600 °C, 800 °C, and 1000 °C) on insoluble organic matter (IOM) extracted from the meteorite Cold Bokkeveld (CM2). For each of the experimental charges, the bulk isotopic compositions of H, N, and C were analyzed using IRMS, the H isotopic heterogeneities (occurrence of hotspots) of the samples were measured by NanoSIMS, and the functional group chemistry and ordering of the IOM was evaluated using FTIR and Raman spectroscopy, respectively. Organic matter in particles heated to ≥ 600 °C during atmospheric entry experienced significant alteration. Loss of isotopically heavy, labile H and N groups results in decreases in bulk $\delta D$, $\delta^{15}N$, H/C and, upon heating ≥800 °C, in N/C. The H heterogeneity was not greatly affected by flash heating to ≤600 °C, although the hotspots tended to be less isotopically anomalous in the 600 °C sample than in the 400 °C sample. However, the hotspots all but disappeared in the 800 °C sample. Loss of C=O groups occurred at 800 °C. Based on the Raman G-band characteristics, the heating resulted in increased ordering of the polyaromatic component of the IOM.

The data presented in this study show that all aspects of the composition of organic matter in IDPs are affected by atmospheric entry heating. Modelling and temperature estimates from stepwise release of He has shown that most IDPs are heated to >500°C (Love and Brownlee, 1991; Nier and Schlutter, 1993; Joswiak et al., 2007), hence, atmospheric entry heating is expected to have altered the organic matter in most IDPs.

Interplanetary dust particles, atmospheric heating, hydrogen, nitrogen, carbon, insoluble organic matter

# 1 Introduction

Organic matter is an important component in many solar system objects such as carbonaceous chondrites, interplanetary dust particles (IDPs) and comets, contributing up to several weight percent of the total mass of primitive chondritic meteorites (Alexander et al., 2007) and almost 50 % of the mass in the dust fraction of at least some comets (Bardyn et al., 2017). The organic matter in meteorites include the so-called building blocks of life, e.g., amino acids and nucleobases, and it has been hypothesizes that the delivery of organic matter by meteorites, comets, and IDPs could have played an important role in the development of life on Earth (Ehrenfreund and Charnley, 2000; Chyba and Sagan, 1992; Anders 1989; Oró 1961). Studies of the most pristine organic matter found, e.g., in IDPs and CR chondrites, has greatly improved our understanding of the nature of this organic matter (for recent reviews see Alexander et al. 2017 and Galvin et al. 2018). Still, many questions remain about the formation and evolution of this organic matter, including how, where and when it formed. One of the most prominent features of primitive organic matter is its high D/H and $^{15}N/^{14}N$ ratios compared to terrestrial materials and that these isotopic compositions tend to be heterogeneous. Insoluble organic matter (IOM) is the dominant organic material in chondrites, as well as the dominant carrier of C, N and noble gases, and similar material appears to be present in IDPs and comets (Alexander et al. 2017; Fray et al. 2017). Bulk IOM samples have $\delta D$ values up ~3500 ‰ and $\delta^{15}N$ values up to ~400 ‰ (Alexander et al., 2007), while so-called hotspots within IDPs and meteoritic IOM can have δD values of up to ~50,000 ‰ and $δ^{15}N$ of ~3000 ‰ (Busemann et al., 2006; Messenger, 2000; Floss et al., 2004).

All meteorites have experienced some degree of parent body modification (Zolensky et al., 2018) while some IDPs, the anhydrous, chondritic porous (CP) IDPs, appear to largely have escaped parent body modifications (Bradley, 2014). However, all IDPs have experienced some degree of atmospheric entry heating. Depending on their speed, angle of entry, density and size, dust particles will either evaporate, melt, or remain un-molten but heated to a few hundred degrees centigrade during atmospheric entry (Love and Brownlee, 1991). Severely heated but un-molten particles show features such as magnetite rims, transformation of the organic matter to a more ordered structure, and formation of vesicular C during volatile escape (Keller et al., 2004; Chan et al., 2019). Observations of nuclear tracks in IDPs provide an upper limit to the heating experienced by some IDPs of 600-650 °C (Fraundorf et al., 1982; Bradley et al., 1984), and stepped heating release of He from IDPs has been used to infer peak atmospheric entry temperatures of 550-1000 °C (Nier and Schlutter, 1993; Joswiak et al., 2007). To be able to compare the organic matter in IDPs with that in chondrites and comets, as well as the nature of the material that would have been delivered to the surface of the early Earth, it is essential to understand how atmospheric entry heating alters IDP organic matter.

While some IDPs contain isotopically very anomalous organic matter, the bulk D/H ratios of IDPs are mostly lower than the D/H ratios of the most isotopically anomalous IOM extracted from CR chondrites, Lewis Cliffs (LEW) 85332 (C3-ung), and Bells (C2-ung) (Alexander et al., 2017, see their Fig. 10). This could be due to atmospheric heating as D-rich functional groups tend to be lost more easily from meteoritic organic material during heating than less anomalous material (e.g., Keller et al., 2004; Okumura and Mimura, 2011; Herd et al., 2011; Remusat et al., 2019). Compared to IOM from Murchison, both anhydrous and hydrous IDPs have higher aliphatic $CH_2/CH_3$ ratios indicating that the aliphatic chains in Murchison are shorter and/or more branched than those in the IDPs (Flynn et al., 2003). This difference could be a primary generic difference between organic matter in chondrites and IDPs, or it could be due to secondary processing of organic matter in meteorites, IDPs, or both. In meteorites, secondary processing is due to thermal or aqueous alteration, and in IDPs secondary processing could occur during irradiation in space or during atmospheric entry (Alexander et al., 2017).

In this study, we conducted a series of flash heating experiments of IOM from Cold Bokkeveld (CM2) to evaluate the effects of atmospheric entry heating on organic matter in IDPs. IOM is the major component of organic matter in primitive meteorites; more than 75 wt. % of the organic matter in meteorites is IOM and there are many similarities between IOM in meteorites and organic matter in IDPs, making it a good proxy for organic matter in IDPs (Alexander et al., 2017). The IOM was extracted from the interior of Cold Bokkeveld and not affected by atmospheric entry heating. These effects are restricted to a few mm from the fusion crust of a meteorite due to the short duration of the heating and poor thermal conductivity of rocky materials. In contrast to the CP IDPs, Cold Bokkeveld has seen parent body alteration, which will be considered in the evaluation of the data.

## 2 Material and Methods

### 2.1 IOM flash heating

IOM was extracted from Cold Bokkeveld (CM2) by the method described in Alexander et al. (2007). Small amounts of IOM (typically ~0.5 mg at a time) were flash heated in Ar-flow to peak temperatures of 400 °C, 600 °C, 800 °C, and 1000 °C using a CDS 1000 pyroprobe at the Geophysical Laboratory. The temperature was increased to the peak temperature with a ramp rate of 500 °C/s, and the samples were kept at the peak temperature for 4 s to simulate atmospheric entry conditions (c.f., Love and Brownlee, 1991).

### 2.2 IRMS

The samples were analyzed for H, C, and N elemental and isotopic compositions at the Geophysical Laboratory. For C and N analyses, we used a Thermo Scientific Delta V$^{Plus}$ mass spectrometer interfaced to a Carlo Erba (NA 2500) elemental analyzer via a Conflo III interface. Hydrogen analyses were conducted with a Thermo Finnigan Delta$^{Plus}$ XL mass spectrometer connected to a Thermo Finnigan Thermal Conversion elemental analyzer (TC/EA) operating at 1400 °C. $N_2$ and $CO_2$ references gases were introduced via the Conflo III, while a dual inlet system facilitated the use of a $H_2$ reference gas of known δD composition (-123.39 ‰ SMOW). Internal working gas standards were analyzed at regular intervals during an analysis to monitor the internal precision of the measured isotopic ratios and elemental compositions throughout the run. House standards, which included both liquid and solid materials, were also analyzed at regular intervals between samples to calibrate and correct the data. The house standards have been calibrated against international (SMOW, NBS-22, air) and commercially certified standards from Isoanalytical, USGS, NBS and Oztech. An $H_3^+$ correction was determined and applied to the H measurements (Sessions et al., 2001). The reported uncertainties for the elemental and isotopic analyses correspond to 1σ compositional variability of internal standards.

Samples were weighed into foil capsules (Ag for H, Sn for C and N) and then stored in a desiccator. Typical solid sample sizes were 0.15 – 0.35 mg. Prior to H analysis, the samples were transferred to a zero-blank autosampler and flushed with He for at least 1 hour to minimize the amount of water absorbed from the atmosphere (Alexander et al., 2010). Blanks were run between different samples to reduce the memory effects. Furthermore, memory effects were monitored by analyzing house standards of H-bearing organic residue solids (e.g., syn-IOM; 67-1183 ‰). There is no memory effect for the C and N analyses.

2.3 Raman

Raman spectra were collected on IOM fragments deposited on a glass slide with a Witec α- SNOM (Scanning Near Field Optical Microscope) with a 532 nm laser connected to a confocal imaging Raman spectrometer at the Geophysical Laboratory. The data were collected in mapping mode using 1-2 s integration times and a laser power of ~ 55 µW. Scan sizes were typically between 7×7 and 20×20 µm, depending on particle size, with a resolution of 3 px/µm. For each sample, 4-7 areas were analyzed, resulting in ~3500-9700 spectra per sample. The spectra were fitted using the procedure described in Busemann et al. (2007); each spectra was fitted to Lorentzian profiles in the area 850-2100 cm$^{-1}$ with a floating linear background. Poor quality spectra were excluded (see Busemann et al., 2007) and the D and G band parameters of the remaining ~2000 spectra/sample were averaged. Light from a Ne lamp was used during the measurements as an external peak position calibration. The peak offsets follow a 3rd order polynomial function with a large variation in the correction for wavenumbers <1200 (Liu et al., 2017). However, the fit is essentially flat in the area of the D and G bands (~1350-1580 cm$^{-1}$) and an average offset calculated from 6 Ne-peaks between 1710 cm$^{-1}$ and 2393 cm$^{-1}$ was used. The offset of 2.32 ± 0.41 cm$^{-1}$ was constant over the three days of Raman measurements.

2.4 NanoSIMS

Unheated and heated IOM fragments were embedded in S and microtomed to a thickness of about 1 μm with a Leica ultramicrotome equipped with a diamond knife. The slices were placed on clean Au-foil mounted onto an Al-stub and the S was subsequently evaporated by heating the Al-stub to 80-100 °C. Hydrogen isotope ion maps of the IOM were collected using the NanoSIMS 50L at the Department of Terrestrial Magnetism. A probe current of ~3 pA, resulting in a spatial resolution of ~170 nm, was used to analyze areas with raster sizes of 20 × 20 μm, 256 × 256 pixels in 10 frames (repeats of each image). Prior to analysis, an area of 30 × 30 μm was presputtered by a higher beam current to reduce surface contamination and implant Cs. The fragments were simultaneously mapped for H, D, $^{13}$C, $^{16}$O, and secondary electrons.

The images were processed using the L'IMAGE software (L. R. Nittler, Carnegie Institution, http://limagesoftware.net). The images were corrected for the 44 ns deadtime of the electron multiplier counting system and the frames of the images were automatically aligned to correct for any beam drift. The area covered by IOM was determined for each image: an ROI (region of interest) including all material with a $^{13}$C count rate ≥10 % of the maximum $^{13}$C count rate was created. For each sample, the D/H of the ROIs and the bulk D/H determined using IRMS were used to determine the instrumental mass fractionation (IMF). After IMF correction, each ion image was automatically segmented into hexagon shaped ROIs with diameters of 781 nm to improve counting statistics at the expense of spatial resolution (e.g., each ROI is much bigger than the intrinsic spatial resolution of the images set by the primary ion beam size of ~170 nm). Following Busemann et al. (2006), a ROI was considered anomalous if $\delta D_{ROI} - \delta D_{unheated} > 3\sigma$ ROI and $1\sigma$ ROI <25‰, where $\sigma$ is the one standard deviation in the $\delta D$ of the ROI. Heterogeneity is measured as the area percentage of isotopically anomalous ROIs divided by the total analyzed area.

2.5 MicroFTIR

Additional microtomed slices of the unheated and heated IOM were placed on BaF$_2$ windows, and the S was evaporated from the window by heating to ~80 °C. The samples were prepared right before analysis to avoid potential surface contamination (Kebukawa et al., 2009) and analyzed with a Jasco FT/IR-6300 +IMV4000 Fourier transform IR microspectrometer at the Geophysical Laboratory using a Dual MCT detector. The spectra were collected in the range 7800-750 cm$^{-1}$ with a resolution of 4.0 cm$^{-1}$ and an accumulation of 4096 scans of IR transmission spectra.

For each sample, between two and five slices provided a strong enough signal to detect the spectral features. For the unheated and 800 °C samples, all slices were extracted from different IOM grains. The two spectra of the 400 °C and 1000 °C samples are from slices of one grain each, and the three 600 °C spectra are from slices from two different grains. A background spectrum was collected through the $BaF_2$ window before each analysis and subtracted from the IOM spectrum.

The spectra contain two regions of interest: the aliphatic stretching region at 3100-2750 cm$^{-1}$, and the 1900-750 cm$^{-1}$ region that includes the carbonyl/ester C=O stretching modes, aromatic C=C stretching modes, and a complex region of fine vibrational modes with wavenumbers below ~1500 cm$^{-1}$. A linear baseline was applied to the each of these regions and the spectra were normalized so that the absorbance at 1600 cm$^{-1}$ = 1 (c.f., Kebukawa et al., 2011; Querico et al., 2018; Remusat et al., 2019). Following Quirico et al. (2018), three parameters were extracted from the spectra: (1) the abundance of carbonyl (C=O) relative to C=C given by the normalized intensity of the peak at ~ 1710 cm$^{-1}$, (2) the aliphatic $CH_2/CH_3$ ratio from the intensities of the antisymmetric stretching bands at ~2925 cm$^{-1}$ and 2955 cm$^{-1}$, and (3) the aliphatic abundance ("ali") relative to C=C given by the integrated absorbance of the normalized aliphatic stretching bands in the range 3000-2800 cm$^{-1}$. In some spectra, a broad feature in the 3600-3200 cm$^{-1}$ region, likely due to O-H stretching modes (Kebukawa et al., 2011), was detected. However, as the feature was faint and the baseline subtraction challenging, we could not assess the evolution of this feature between the samples.

# 3 Results

## 3.1 IRMS

The IRMS data show that in general, the IOM develops lighter δD and $δ^{15}N$ compositions and lower H/C and N/C ratios with increasing temperature (Fig. 1, Table 1). The largest effect on the isotopic composition is seen in H: the **δD of the 400 °C** sample is ~300 ‰ more enriched than that of the 1000 °C sample (Fig. 1A). The decrease of the H/C ratio with increasing heating from 54 ± 4 (at. × 100) in the 400 °C sample to 27 ± 2 (at. × 100) in the 1000 °C shows that the **change in δD is coupled** to the H loss from the IOM. One recent study has reported the opposite, i.e., that H/C in IDPs increase with heating (Chan et al., 2019). However, this effect disappears when one only compares C-rich areas in the IDPs, indicating that the effect could be due to contamination or possibly the presence of undetected hydrogen-bearing minerals.

The effect on the N isotopic composition is also clearly resolvable; $δ^{15}N$ is ~15 ‰ more enriched in the 400 °C sample than in the 1000 °C sample (Fig. 1C). This is accompanied by a modest decrease in the N/C ratio from 3.53 ± 0.05 (at. × 100) in the 400 °C sample to 3.31 ± 0.05 (at. × 100) in the 1000 °C sample. The change to the $δ^{13}C$ compositions is much smaller, with a <1‰ difference between the 400 °C and the 1000 °C samples (Fig. 1B), similar to the decrease in δ¹³C seen in previous heating experiments (Mimura et al., 2007; Okumura and Mimura, 2011). The 400 °C sample has more enriched δD and δ¹⁵N values than the unheated IOM, likely due to terrestrial contamination of the unheated sample or loss of functional groups that isotopically exchanged during IOM isolation.

Table 1. H, N, C elemental and isotopic composition of the samples as measured by IRMS. The atomic ratios have been multiplied by 100.

| Sample | δD (‰) | $δ^{15}N$ (‰) | $δ^{13}C$ (‰) | H/C (atom) | N/C (atom) |
|---|---|---|---|---|---|
| Unheated | 725 ± 6 | -1.8 ± 0.2 | -18.0 ± 0.2 | 54 ± 4 | 3.53 ± 0.05 |
| 400°C | 871 ± 3 | -0.3 ± 0.2 | -18.1 ± 0.2 | 55 ± 6 | 3.53 ± 0.05 |
| 600°C | 661 ± 2 | -5.6 ± 0.2 | -18.4 ± 0.2 | 35 ± 2 | 3.49 ± 0.05 |
| 800°C | 604 ± 6 | -12.3 ± 0.2 | -18.9 ± 0.2 | 28 ± 2 | 3.41 ± 0.05 |
| 1000°C | 567 ± 6 | -15.4 ± 0.2 | -18.8 ± 0.2 | 27 ± 2 | 3.31 ± 0.05 |

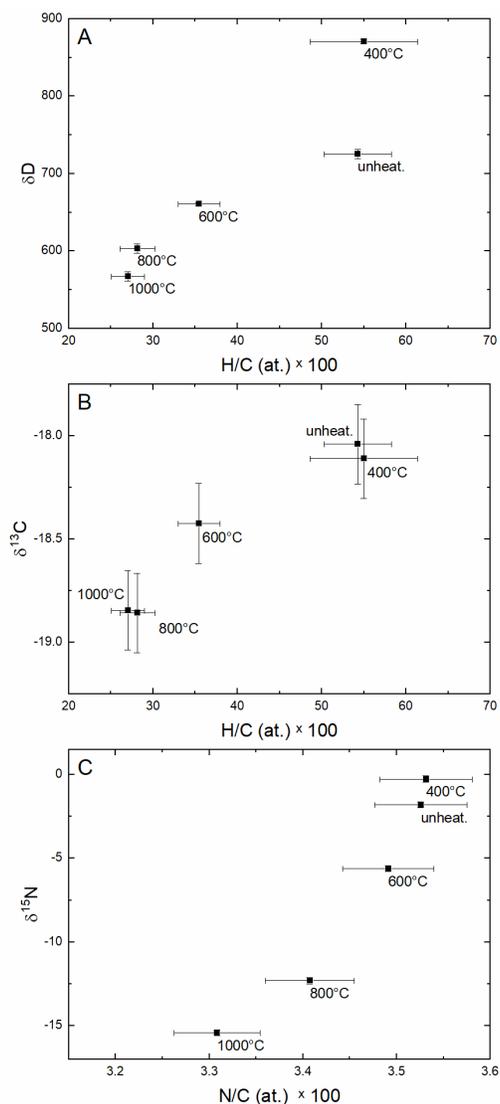

Figure 1. Isotopic and elemental compositions of the unheated and heated IOM samples: (A) the δD and H/C values of the samples, (B) the δ¹³C and H/C values, and (C) the δ¹⁵N and N/C values. In general, with increased heating the IOM became isotopically lighter and H/C and N/C decreased. The 400 °C step does not follow this trend but instead has higher δD and δ¹⁵N values than the unheated sample, probably due to the loss of terrestrial contamination with normal isotopic compositions during the 400 °C heating step.

3.2 Raman

The Raman spectra of IOM are dominated by two bands: the D-band at ~ 1360 cm$^{-1}$ and the G-band at ~ 1586 cm$^{-1}$. The G-band (G for graphite) is due to in-plane stretching of neighboring sp$^2$ atoms and only reflects the graphitic ordering. The D-band (D for disorder) is a mixture of different resonances and reflects defects, crystal boundary effects, as well as defects due to heteroatoms such as O (Beyssac et al., 2002; Pasteris and Chou, 1998).

The heating systematically changed the G-band characteristics of the IOM so that the peak became narrower and moved towards higher wavenumbers. The G-band peak width decreased from 100.6 ± 0.7 cm$^{-1}$ in the unheated sample to 88.4 ± 0.7 cm$^{-1}$ in the 1000 °C sample, and the G-band peak position moved from 1583.5 ± 1.4 cm$^{-1}$ in the unheated sample to 1589.2 ± 2.3 cm$^{-1}$ in the 1000 °C sample (Fig. 2A, Table 2). No changes were observed in the D-band characteristics (Table 2), likely reflecting that the G-band provides a clearer signal than the D-band. This has been attributed to the lesser effects of the particle orientation with respect to the incident laser beam on the shape of G-band (Foustoukos, 2012; Jenniskens et al., 2009; Ross et al., 2011). The shift in the G-band peak position has been interpreted as the development of nanocrystalline graphite during which the D' band appears at ~1620 cm$^{-1}$ (Ferrari and Robertson 2000). The D' band is not resolved from the G-band in small grains and results in an apparent shift in the G-band peak position (Ferrari and Robertson 2000).

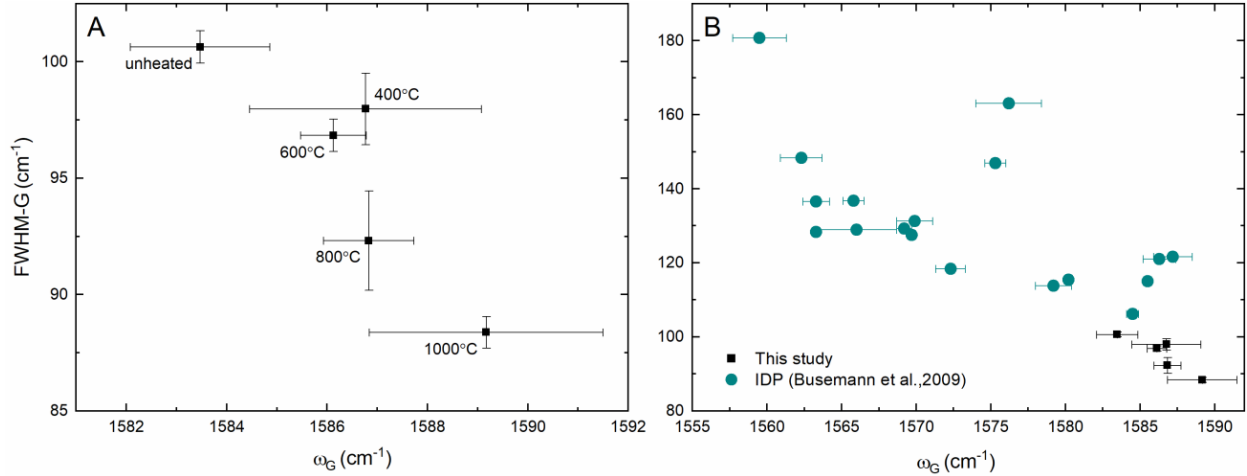

Figure 2. The Raman G-band parameters for the Cold Bokkeveld IOM heated to different temperatures. The G-band peak width (FWHM) becomes narrower with increased heating and there is a tendency for the G-band peak position (ω) to move towards higher wavenumbers. (B) The data in this study compared to the range of G-band parameters determined for IDPs by Busemann et al. (2009).

Table 2. Raman parameters of the samples. Uncertainties are standard deviations of all used spectra. Peak position uncertainties include uncertainty of the external Ne light correction.

| Sample | total #spectra | used #spectra | D peak position | D FWHM | G peak position | G FWHM |
|---|---|---|---|---|---|---|
| Unheated | 4329 | 2351 | 1358.6 ± 1.3 | 304 ± 2 | 1583.5 ± 1.4 | 100.6 ± 0.7 |
| 400°C | 3519 | 2282 | 1360.8 ± 1.6 | 290 ± 8 | 1586.8 ± 2.3 | 98.0 ± 1.5 |
| 600°C | 5292 | 2331 | 1360.7 ± 0.6 | 302 ± 2 | 1586.1 ± 0.7 | 96.8 ± 0.7 |
| 800°C | 4572 | 1663 | 1359.3 ± 1.1 | 302 ± 4 | 1586.8 ± 0.9 | 92.3 ± 2.1 |
| 1000°C | 9720 | 3916 | 1358.7 ± 2.0 | 293 ± 12 | 1589.2 ± 2.3 | 88.4 ± 0.7 |

## 3.3 NanoSIMS

The NanoSIMS data provide information on the spatial heterogeneity of the H isotopic compositions of the IOM samples that are primarily due to the abundances of isotopic hotspots. The data were collected to determine if and at what temperature these hotspots disappear during flash heating. Hydrogen isotopic maps of representative areas of the IOM samples show that the hotspots survive flash heating to 600 °C but not to 800 °C (Fig. 3). This visual observation was confirmed by the heterogeneity calculation (section 2.4: a ROI was considered anomalous if $\delta D_{ROI} - \delta D_{unheated} > 3\sigma$ ROI and $1\sigma$ ROI <25%, where $\sigma$ is the one standard deviation in the $\delta D$ of the ROI). The unheated, 400 °C and 600 °C samples were determined to have spatial heterogeneities of 3.4-3.6 area% (Table 3), similar to the Murchison (CM2) value of 4.3 area% reported by Busemann et al. (2006). The heterogeneity drastically decreased in the 800 °C sample to 0.4 area%, and in the 1000 °C sample to 0.1 area%.

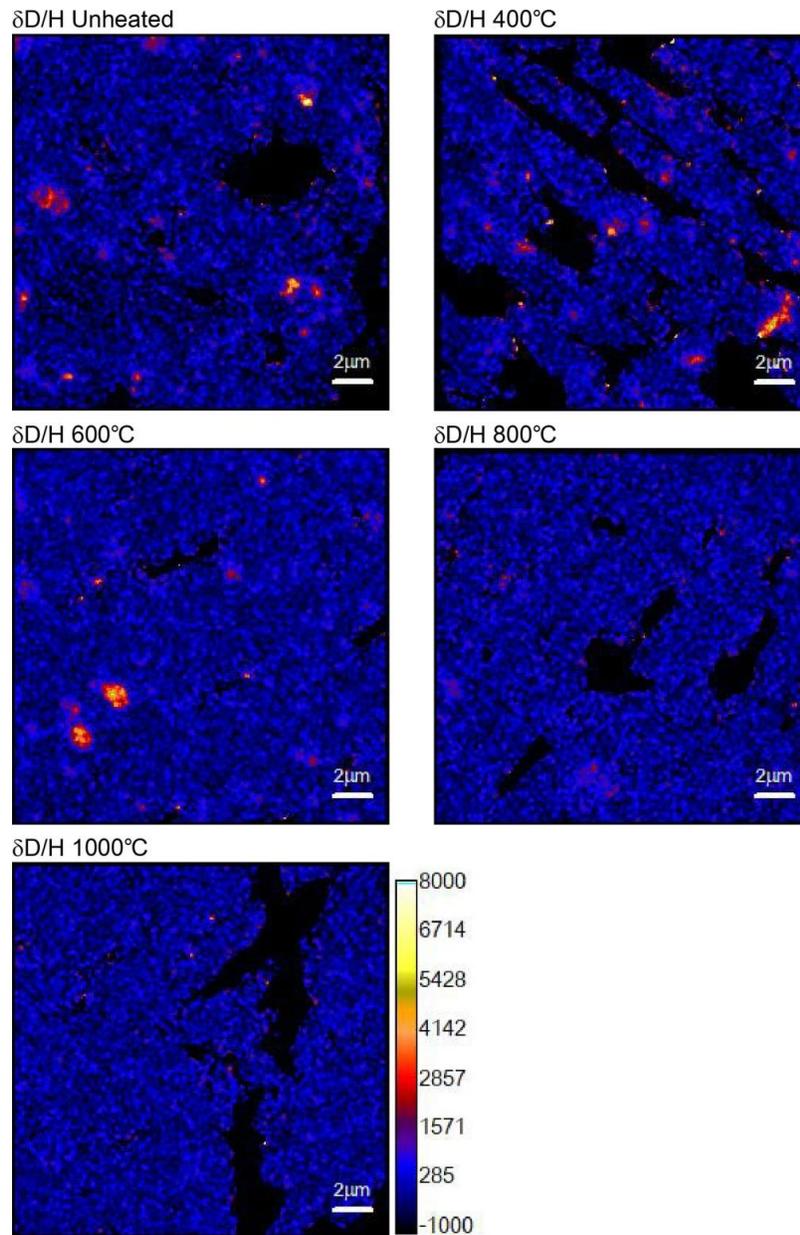

Figure 3. Representative NanoSIMS δD ion images of the different samples. Hotspots are clearly visible in the samples heated up to 600 °C, but are essentially gone in the sample heated to 800 °C and above.

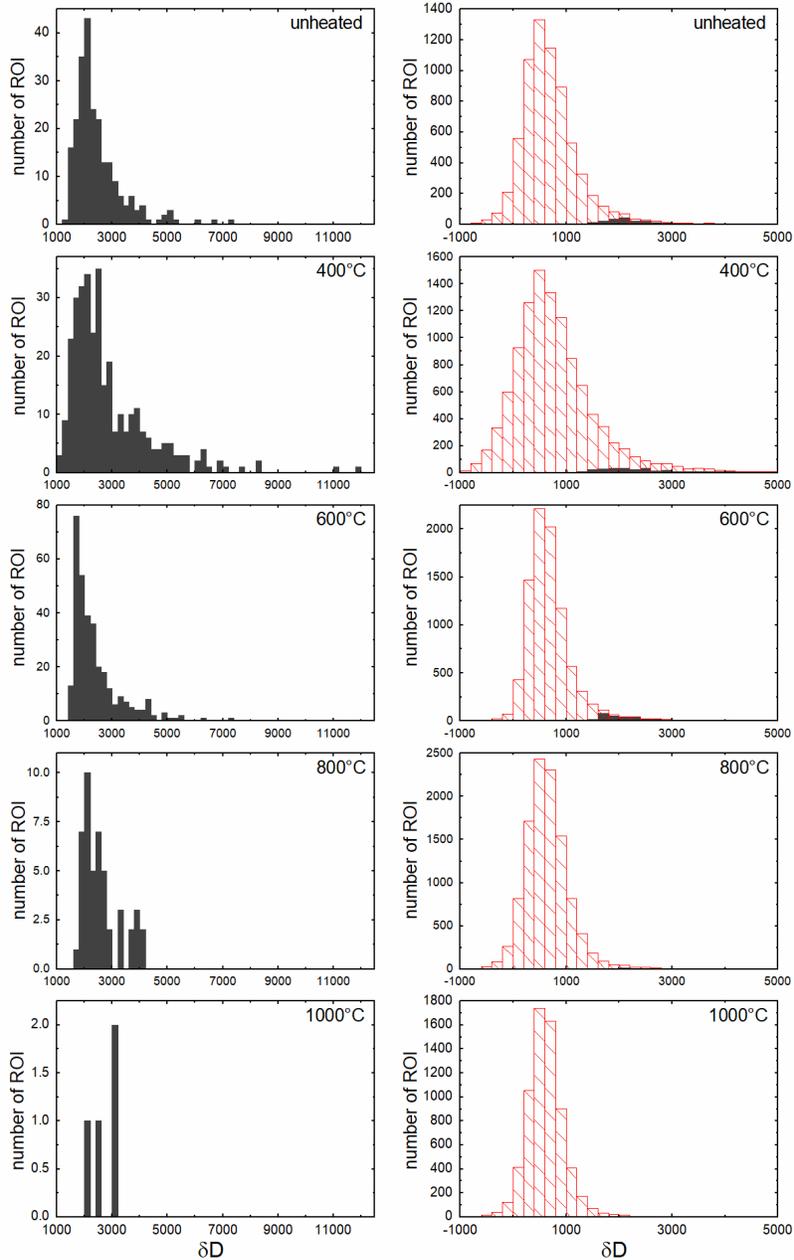

Figure 4. Histograms showing the distribution of δD values in anomalous ROIs (left) and of all ROIs as well as anomalous ROIs (right) for the different samples. The distribution of all ROIs, as well as those of the anomalous ROIs, become narrower with increasing peak heating temperature.

The histograms in Fig. 4 show the distribution of δD values in the anomalous ROIs (left column) and all the ROIs (right column). The distribution of the δD values of all the ROIs is narrower at 600 °C, 800 °C and 1000 °C than in the unheated and 400°C samples. The distribution of δD values amongst anomalous ROIs also appears to be narrower in the 600 °C sample than in the 400 °C sample. It therefore appears that, while there is no decrease in heterogeneity, as defined here, in these two samples, the hotspots became less anomalous with increasing temperature. Presumably, some of the more D-rich functional groups were lost as a step towards full destruction of the hotspots.

Table 3. Total area mapped for δD using NanoSIMS and the heterogeneity determined for each sample.

| Sample | Analyzed area [µm$^2$] | Heterogeneity[1] |
|---|---|---|
| Unheated | 2818 | 3.4% |
| 400 °C | 3935 | 3.4% |
| 600 °C | 3681 | 3.6% |
| 800 °C | 4358 | 0.4% |
| 1000 °C | 2672 | 0.1% |

[1] Heterogeneity was determined following Busemann et al. (2006). Each ion image was automatically divided into hexagon shaped ROIs with diameters of 781 nm using L'IMAGE (Larry R. Nittler). A ROI was considered anomalous if $\delta D_{ROI} - \delta D_{average} > 3\sigma_{ROI}$ and $\sigma_{ROI}$ <25%. Heterogeneity is presented as the percentage of anomalous area divided by the total analyzed area.

## 3.4 MicroFTIR

Using the FTIR spectra, we can constrain the evolution of molecular structure in the IOM during flash heating. We identified the peaks of the functional groups based on the list of peaks in IOM FTIR spectra and possible assignments provided by Kebukawa et al. (2011; their Table 2).

Figure 5 shows the FTIR spectra in the aliphatic C-H stretching region, 3050-2800 cm$^{-1}$ (Fig. 5A), and the 1900-750 cm$^{-1}$ region that includes the carbonyl/ester C=O stretching modes, aromatic C=C stretching modes, and a complex region of fine vibrational modes with cm$^{-1}$ lower than ~1500 (Fig. 5B). The most obvious change to the FTIR spectra with flash heating is the decrease in the intensity of the C=O band at ~1720 cm$^{-1}$, relative to the C=C band, in the samples heated to 800 °C or more (Fig. 5B, 6A). The C=O band is likely due to carboxyl groups (COOH), thus the loss is an indication of thermal decarboxylation. A second difference is that the aromatic C=C stretching band moved towards lower frequency in the 800 °C and 1000 °C samples compared to the unheated, 400 °C, and 600 °C samples (Fig. 5B). Kebukawa et al. (2011) observed the same difference in the aromatic C=C band between IOM from meteorites that had been altered to different degrees, interpreted as altering of the IOM to more condensed aromatic structures during thermal metamorphism. A third observation is that the $CH_2/CH_3$ ratio is higher and the ali parameter lower in the 800 °C sample than in the other samples (Fig. 6B). While the $CH_2/CH_3$ ratio is slightly higher and the ali parameter slightly lower in the 1000 °C samples than in the unheated, 400 °C, and 600 °C samples, these differences are considerably smaller than for the 800 °C sample.

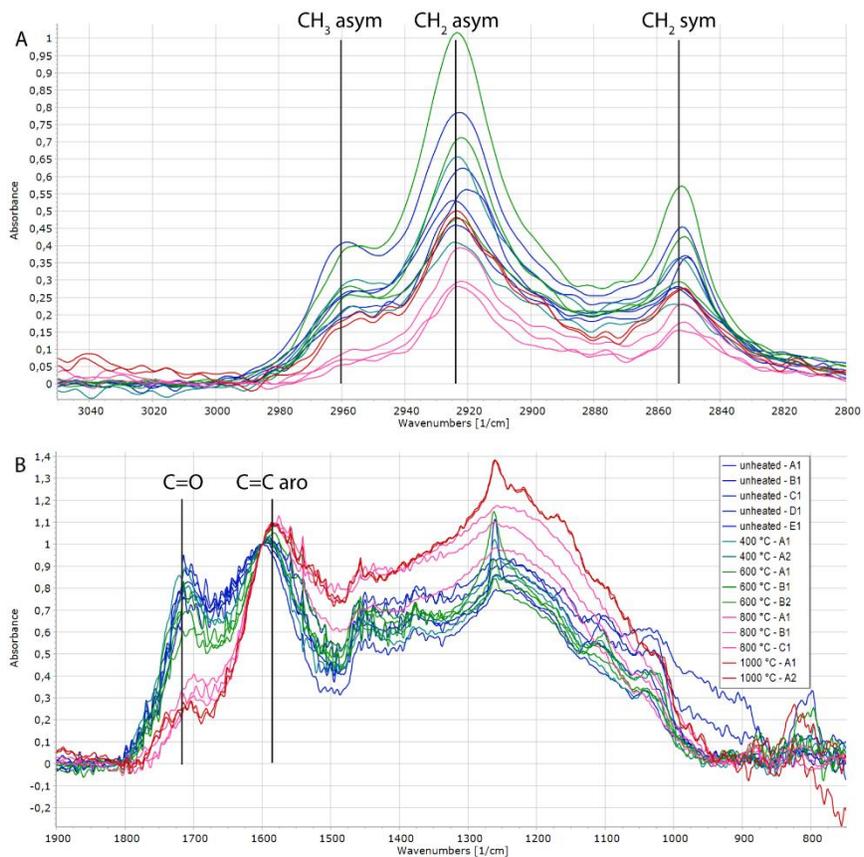

Figure 5. FTIR spectra of the samples normalized to an absorbance of 1 at 1600 cm$^{-1}$, the C=C stretch peak position in the unheated sample. (A) The aliphatic stretching region 3050-2800 cm$^{-1}$. (B) The spectra in region 1900-750 cm$^{-1}$. The relative decrease in the intensity of the C=O peak is clearly seen.

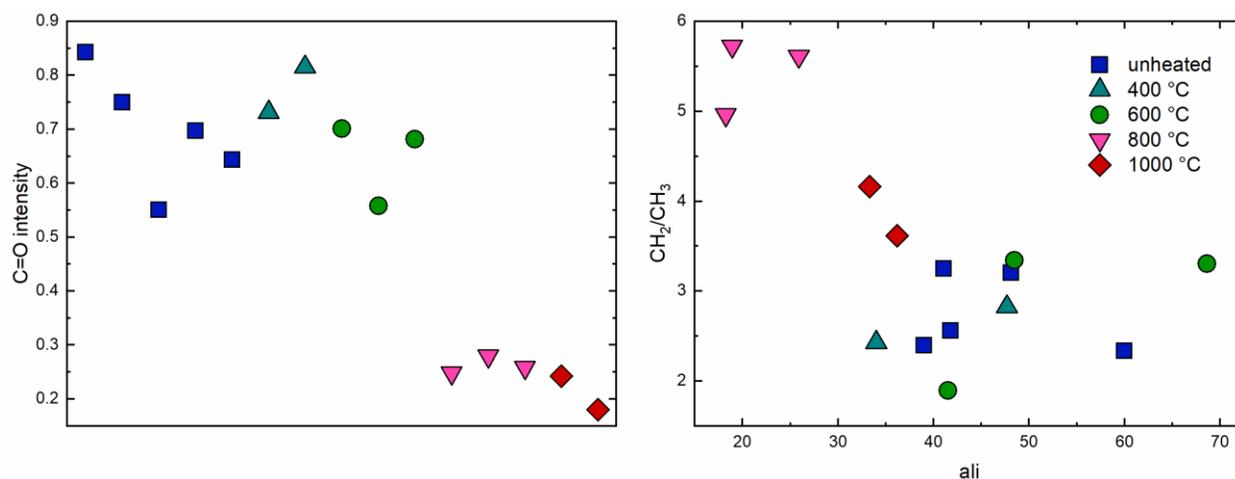

Figure 6. (A) The intensity of the C=O peak (~1710 cm$^{-1}$) in individual spectra normalized so that the intensity at 1600 cm$^{-1}$ has a value of 1. There is a clear drop in the intensity of the C=O peak in the samples with peak temperatures of 800 °C and 1000 °C compared to the unheated sample and the samples with peak temperatures of 400 °C and 600 °C. (B) The intensity ratio of the asymmetric $CH_2/CH_3$ stretching plotted against the integrated area of the aliphatic region "ali". The 800 °C sample as a lower ali value and higher $CH_2/CH_3$ ratios than the other samples while the 1000 °C sample plots closer to the unheated, 400 °C and 600 ° samples.

## 4 Discussion

### 4.1 Effects of flash heating on IOM

Compared to the 400 °C sample, the δD value of the bulk IOM samples decreased by ~ 200 ‰ during heating to 600 °C and by ~ 300 ‰ during heating to 1000 °C (Fig. 1, Table 1). A smaller decrease in δD of ~100‰ when heated to 800 °C was observed in pyrolysis residues of Murchison IOM (Okumura and Mimura, 2011). Shock experiments with a peak pressure of 7.3

GPa and a peak temperature of 1030 °C resulted in a larger decrease in the δD value than our samples of ~500‰ (Mimura et al., 2007). It is well known that H in IOM tends to become isotopically lighter upon heating (e.g., Herd et al., 2011; Keller et al., 2004; Remusat et al. 2009; Okumura and Mimura, 2011). This feature likely reflects that thermally labile aliphatic compounds are more D-rich than more heat resistant compounds in the IOM (Keller et al., 2004; Remusat et al., 2009; Mimura et al., 2007; Okumura and Mimura, 2011). Interestingly, the δD heterogeneity of the 600 °C sample of 3.6 ‰ is similar to that of the 400 °C and unheated samples (Table 3), demonstrating that the decrease in bulk δD during heating to 600 °C may not be coupled with loss of hotspots. However, the anomalous ROIs in the 400 °C sample have a larger spread towards higher δD value than the ROIs in the 600 °C sample (Fig. 4), indicating that some D was lost from the hotspots.

Our flash heating results stand in contrast to Remusat et al. (2009) who found that all hotspots in Murchison (CM2) IOM disappeared after pyrolyzing at 600°C. However, there is very little information on the experimental conditions and it is possible that their sample was heated for considerably longer time than our samples. Remusat et al. (2019) found that heating of IOM from Orgueil (CI) for 1h resulted in complete loss of δD hotspots. The retention of hotspots in IOM from Cold Bokkeveld during flash heating to 600 °C compared to destruction of hotspots in IOM from Orgueil during 1h of heating at 500 °C probably reflects the kinetics of hotspot destruction. This highlights the importance of conducting flash heating experiments to simulate atmospheric entry heating rather than extrapolating the results from longer heating experiments.

Considering such possible kinetic effects, had the reaction been allowed to continue for significantly longer than 4s in our heating experiments, then the hotspots would probably have been lost from the 600 °C sample. The loss of D-rich material is, however, fast enough during flash heating to 800 °C to almost completely lose the D-hotspots and during flash heating to 1000° C fast enough to fully homogenize the IOM (Fig. 3, Table 1).

The $\delta^{15}N$ also decreased with increased heating, although the effect was significantly smaller than that of δD (Fig.1; Table 1). This is in agreement with previous studies showing that the labile component of IOM has higher $\delta^{15}N$ than the more refractory component (Alexander et al., 1998; Sephton et al., 2003). Lowering of the $\delta^{15}N$ to the same magnitude as in this study was observed in flash heating of some bulk Orgueil (CI) samples to 1350 °C for ≥ 7s (Füri et al., 2013). Okumura and Mimura (2011) detected mainly pyridine, pyrrole and their derivatives (pyridine group) in pyrolysates from Murchison IOM heated to 450 °C and 550 °C, and nitrile and alkyl derivatives (nitrile group) in 650 °C and 800 °C steps. Since most of the nitrile group compounds are acid-hydrolysable and lost during IOM preparation, Okumura and Mimura concluded that these were likely secondary products from N heterocycles. In addition to these groups, Okumura and Mimura (2011) deduced, based on the N mass balance, that small N-bearing groups were released due to dissociation of N-bearing linkages and substitutes such as nitro and amino groups. Some of these compounds are likely the carrier of the thermally label $\delta^{15}N$ enrichment in Cold Bokkeveld IOM, but with our current data set we are not able to determine which. Keller et al. (2004) argued based on observations in IDPs that the carrier of the $\delta^{15}$N enrichment was of amine functionality and likely attached to aromatic hydrocarbons.

The Raman D-and G-bands reflect the structural order of IOM polyaromatic C and are largely controlled by the size and ordering of polyaromatic units and the chemistry of the IOM (Quirico et al., 2009). The G-band parameters changed with increasing heating so that the G-band became narrower and the peak position moved to higher cm$^{-1}$ (Fig. 2A). The Raman parameters change in the same way with maturation of coal (Quirico et al., 2005; Bower et al., 2013), and IOM from meteorites exhibit similar modification due to parent body alteration (Busemann et al., 2007; Quirico et al., 2018). The relatively small changes to the Raman parameters in the samples in this study (G-band FWHM 88-101 cm$^{-1}$) compared to those in natural samples (G-band FWHM 53-109 cm$^{-1}$ among carbonaceous chondrites; Busemann et al., 2007) is almost certainly a kinetic effect. Cody et al. (2008) showed that there is a well-resolved time-temperature relationship for the development of the 1s-$\sigma$* exciton in C-XANES IOM spectra (their Fig. 5). They further showed that there is a correlation between Raman parameters and the 1s-$\sigma$* exciton in IOM, demonstrating that the changes in Raman parameters are also controlled by modification kinetics and temperature. If translated to extended parent body alteration (10$^7$ years), the difference in the G-band that occurred between our unheated and 1000 °C samples correspond only to a temperature difference of ~115 °C (Cody et al., 2008). Developing a kinetic expression for Raman parameters is outside the scope of this study but would be an interesting focus for future work. The changes in the G-band parameters in the flash heated samples are also small compared to the spread observed in IDPs (Fig. 2B). Starkey et al. (2013) observed that Raman parameters for IDPs tend to fall on the same trends as those for IOM, but that the two sample sets mostly do not overlap and that the Raman parameters of the IDPs indicated more primitive structures (Starkey et al., 2013). Interestingly, Dobrică et al. (2011) did not observe the same pattern in micrometeorites collected from Antarctica. In their study, clearly heated micrometeorites (scoriaceous and fine-grained scoriaceous) plotted to the right of a trend defined by the most primitive micrometeorites and chondrites in a G-band FWHM vs. peak position diagram rather than at the lower end of that trend (see their Fig. 5). Dobrică et al. (2011) interpreted this as a combined effect of oxidation and heating in the atmosphere. A similar observation was done by Chan et al. (2019), their CP-IDPs had higher FWHM G-band than meteorite IOM, resulting in the IDPs and the IOMs describing different trends in G-band FWHM vs peak position diagram. Chang et al. (2019) interpreted their data as being either due to structural differences between organic matter in IDPs and meteorites, or a difference in the maturation process between IOM and IDP samples.

4.2 Implications for IDP studies and the quest to find the most pristine IDP particles

Our data show that the organic matter in the IDPs that have been least affected by atmospheric entry have the following characteristics: **high δD and δ$^{15}$N** values, high H/C and N/C ratios, more **hotspots with high δD** values, a broad Raman G-band with a peak position towards lower cm$^{-1}$, and low C=O/C=C$_{arom.}$ ratio. Except for the isotopic compositions, which are sample destructive measurements, the clearest trend in the data was the decrease of the C=O/C=C$_{arom.}$ ratio with heating observed in the FTIR data. However, the small IDPs require bright synchrotron light for acquisition of FTIR spectra, excluding FTIR as a routine measurement to detect the least altered IDPs. In addition, it remains to be confirmed that the C=O that is lost from the IOM during flash heating is present in IDPs as it might be added to the IOM during sample preparation.

A more promising method for identifying the least atmospherically heated IDPs might be Raman spectroscopy. The shift in G-band characteristics is, except for loss of terrestrial contaminants, the only observed alteration to IOM during flash heating to 400°C, illustrating the sensitivity of the Raman parameters to flash heating. The Raman parameters could, therefore, potentially be useful indicators of the primitiveness of IDPs. However, it is currently unsure how much of the variability seen in Raman parameters of IDPs was established prior to atmospheric entry. It is possible that the relatively small changes to the Raman spectra during the flash heating experiments are influenced by the alteration that the IOM experienced in the Cold Bokkeveld parent body. This could be addressed in future studies by repeating the experiments in this study using more primitive IOM from CR chondrites. In addition, a single IDP could be flash heated in a controlled environment while collecting Raman information.

The presence of D/H hotspots in IOM and IDPs is often seen as evidence of the primitiveness of the organic matter (e.g., Busemann et al., 2006; Messenger, 2000). However, we have shown here that structure, bulk isotopic composition, and the isotopic composition of the hotspots themselves can be significantly altered by flash heating while the heterogeneity stays constant. The organic matter in IDPs heated to 600°C could contain hotspots and appear pristine, despite having been significantly altered during atmospheric entry. Minerals may work as catalysts for reactions occurring within IOM during heating (Kebukawa et al., 2010), resulting in even larger effects on the organics than the ones described here. The effects of oxidation during atmospheric entry heating are also poorly understood. Magnetite rims are characteristic of more heated IDPs (Keller et al., 2004), but it seems likely that at least some oxidation of organics will take place at lower temperatures than those needed to form magnetite rims. The data presented in this study show that all aspects of the composition of organic matter in IDPs are affected by atmospheric entry heating. Based on modelling and temperature estimates from stepwise release of He, most IDPs are heated to >500°C (Love and Brownlee, 1991; Nier and Schlutter, 1993; Joswiak et al., 2007) and these effects are expected to be present in most IDPs.

Conclusions

Atmospheric entry heating can result in significant alteration to the organic matter already at a peak temperature of 600 °C and catalytic effects of minerals might result in even larger modifications (Kebukawa et al., 2010). This affects a large fraction of the IDP collection since most IDPs have been heated to > 500 °C (Love and Brownlee, 1991; Nier and Schlutter, 1993; Joswiak et al., 2007). IOM in IDPs heated to 600 °C are expected to 1) have lost D- and $^{15}$N rich compounds resulting in lower bulk $\delta D$, $\delta^{15}N$, and H/C values as well as less anomalous D hotspots, 2) have increased structural ordering of the polyaromatic C based on the G-band

Raman parameters, 3) have lost C=O groups, 4) have retained its D hotspots albeit with lower δD values. These modifications will be larger in IDPs that are heated to higher peak temperatures. In addition, samples heated to 800 °C and 1000 °C will lose their D hotspots, lose N to a higher degree than C resulting in a lower N/C ratio, and perhaps have experienced a slight decrease in $\delta^{13}C$. Flash heating to 400 °C only resulted in loss of weakly bound terrestrial contamination and a slight ordering of the polyaromatic component of IOM.

## Acknowledgment

MR would like to thank Brad De Gregorio for instructions on how to use the ultramicrotome. This work was in part supported by the Swiss National Science Foundation (grant number P2EZP2_165234: MR) and by NASA (grant numbers NNX14AG95G: MR, NNX15AH77G: DF, CA, GC).